%
\let\useblackboard=\iftrue
%
%
\newfam\black
\input harvmac.tex
\def\Title#1#2{\rightline{#1}
\ifx\answ\bigans\nopagenumbers\pageno0\vskip1in%
\baselineskip 15pt plus 1pt minus 1pt
\else
\def\listrefs{\footatend\vskip 1in\immediate\closeout\rfile\writestoppt
\baselineskip=14pt\centerline{{\bf References}}\bigskip{\frenchspacing%
\parindent=20pt\escapechar=` \input
refs.tmp\vfill\eject}\nonfrenchspacing}
\pageno1\vskip.8in\fi \centerline{\titlefont #2}\vskip .5in}

\ifx\answ\bigans\def\tcbreak#1{}\else\def\tcbreak#1{\cr&{#1}}\fi
\useblackboard
\message{If you do not have msbm (blackboard bold) fonts,}
\message{change the option at the top of the tex file.}
\font\blackboard=msbm10 scaled \magstep1
\font\blackboards=msbm7
\font\blackboardss=msbm5
\textfont\black=\blackboard
\scriptfont\black=\blackboards
\scriptscriptfont\black=\blackboardss

\else

\fi
%
\def\yboxit#1#2{\vbox{\hrule height #1 \hbox{\vrule width #1
\vbox{#2}\vrule width #1 }\hrule height #1 }}
\def\fillbox#1{\hbox to #1{\vbox to #1{\vfil}\hfil}}
\def\ybox{{\lower 1.3pt \yboxit{0.4pt}{\fillbox{8pt}}\hskip-0.2pt}}
\def\np#1#2#3{Nucl. Phys. {\bf B#1} (#2) #3}
\def\pl#1#2#3{Phys. Lett. {\bf #1B} (#2) #3}

\def\comments#1{}

\def\half{{1\over 2}}
\def\Tr{{{\rm Tr\  }}}
\def\tr{{\rm tr\ }}

\def\II{\relax{I\kern-.07em I}}

\def\IZ{\relax\ifmmode\mathchoice
{\hbox{\cmss Z\kern-.4em Z}}{\hbox{\cmss Z\kern-.4em Z}}
{\lower.9pt\hbox{\cmsss Z\kern-.4em Z}}
{\lower1.2pt\hbox{\cmsss Z\kern-.4em Z}}\else{\cmss Z\kern-.4em
Z}\fi}
\def\IB{\relax{\rm I\kern-.18em B}}
\def\IC{{\relax\hbox{$\inbar\kern-.3em{\rm C}$}}}
\def\ID{\relax{\rm I\kern-.18em D}}
\def\IE{\relax{\rm I\kern-.18em E}}
\def\IF{\relax{\rm I\kern-.18em F}}
\def\IG{\relax\hbox{$\inbar\kern-.3em{\rm G}$}}
\def\IGa{\relax\hbox{${\rm I}\kern-.18em\Gamma$}}
\def\IH{\relax{\rm I\kern-.18em H}}
\def\II{\relax{\rm I\kern-.18em I}}
\def\IK{\relax{\rm I\kern-.18em K}}
\def\IP{\relax{\rm I\kern-.18em P}}

\font\cmss=cmss10 \font\cmsss=cmss10 at 7pt
\def\IR{\relax{\rm I\kern-.18em R}}

\def\Tr{\rm Tr}

\Title{\vbox{\baselineskip12pt\hbox{hep-th/9609161}
\hbox{RU-96-85}}}
{\vbox{
\centerline{Non-trivial Fixed Points of The Renormalization Group}
\centerline{in Six Dimensions}}}
\centerline{Nathan Seiberg}
\smallskip
\smallskip
\centerline{Department of Physics and Astronomy}
\centerline{Rutgers University }
\centerline{Piscataway, NJ 08855-0849}
\centerline{\tt seiberg@physics.rutgers.edu}
\bigskip
\bigskip
\noindent
We start a systematic analysis of supersymmetric field theories in six
dimensions.  We find necessary conditions for the existence of
non-trivial interacting fixed points.  String theory provides us with
examples of such theories.  We conjecture that there are many other
examples.

\Date{September 1996}

\nref\wittenII{E. Witten, ``Some Comments on String Dynamics,''
Contributed to STRINGS 95: Future Perspectives in String Theory, Los
Angeles, CA, 13-18 Mar 1995. hep-th/9507121}%
\nref\strominger{A. Strominger,  ``Open p-branes,'' hep-th/9512059.}%
\nref\ganoha{O. Ganor and A. Hanany, ``Small E(8) Instantons and
Tensionless Noncritical Strings,'' hep-th/9602120.}%
\nref\seiwit{N. Seiberg and E. Witten, ``Comments on String Dynamics in
Six Dimensions,'' hep-th/9603003, \np{471}{1996}{121}.}%
\nref\dlp{M. Duff, H. Lu and C.N. Pope, ``Heterotic Phase
Transitions and Singularities of The Gauge Dyonic String,''
CTP-TAMU-9-96, hep-th/9603037, \pl{378}{1996}{101}.}%
\nref\threedone{N. Seiberg, ``IR Dynamics on Branes and Space-Time
Geometry,'' hep-th/9606017.}%
\nref\fived{N. Seiberg, ``Five Dimensional SUSY Field Theories,
Non-trivial Fixed Points, and String Dynamics,'' hep-th/9608111.}%
\nref\morsei{D.R. Morrison and N. Seiberg, ``Extremal Transitions and
Five-Dimensional Supersymmetric Field Theories,'' hep-th/9609069.}

Recent developments in quantum field theory have led to the discovery
of a large number of non-trivial interacting fixed points of the
renormalization group.  Some of these theories were discovered using
recent advances in string duality \refs{\wittenII - \morsei}.  

On the moduli space of vacua of some these theories there are string
like excitations.  Their tension approaches zero at singular points in
the moduli space.  Therefore, these theories are often referred to as
``tensionless string theories.''  It is our view that they are
conventional interacting local quantum field theories.  This possible
interpretation was first mentioned in \refs{\wittenII, \seiwit} but
became more clear after one of these theories (in three dimensions) was
given a Lagrangian description
\ref\intse{K. Intriligator and N. Seiberg, ``Mirror Symmetry in Three
Dimensional Gauge Theories,'' hep-th/9607207.}
and the renormalization group flows out of the five dimensional theories
appeared consistent with field theory \fived.

It is likely that some of these theories do not arise under
renormalization group flow from a free field theory at short distance.
Therefore, they do not have a continuum Lagrangian description.  It is
an extremely interesting and challenging problem to find a good
presentation of these theories.

Some of these interacting field theories are in five and six dimensions.
Their existence contradicts the lore saying that no such theories exist.
This lore was based on continuum Lagrangian field theory, in other
words, on perturbation theory around a Gaussian fixed point.  Since
these theories can flow to free field theories, we could still attempt
to describe them using a Lagrangian with an irrelevant operator.  (Of
course, such a description cannot be complete.)  Therefore, we suggest
that perhaps a better way to state the lore is that the coefficient of
the irrelevant operator, the coupling constant, must be infinite at the
non-trivial fixed point.

In this note we start a systematic search for such field theories in six
dimensions with the minimal amount of supersymmetry.  The minimal super
Poincare symmetry has three massless representations with low spin: a
hypermultiplet, a vector multiplet and a tensor multiplet.  It is not
known how to write a Lagrangian including the tensor multiplet because
it includes a two form gauge field whose three form field strength is
self-dual.  Therefore, we start with only vector multiplets and
hypermultiplets. 

The most general Lagrangian (with at most two derivatives) including
only vector multiplets and hypermultiplets is parametrized by a gauge
group $G$ and a matter representation $R$.  It typically has a moduli
space of vacua parametrized by the expectation value of the scalars in
the hypermultiplets (Higgs branch).  Unlike the corresponding five and
four dimensional theories there is no Coulomb branch because the vector
multiplets do not include scalar fields.

In the quantum theory we should also consider the anomalies.  To keep
the discussion simple, we limit ourselves to simple gauge groups.  Then,
the anomaly is
\eqn\anomaly{{\Tr}_a F^4 - {\Tr}_R F^4 = \alpha {\tr} F^4 + {c \over
d^2} ({\tr} F^2)^2 } 
where ${\Tr}_a$, ${\Tr}_R$ and $\tr$ are traces in the adjoint
representation, in the representation $R$ and in the fundamental
representation respectively and $d$ is the dimension of the fundamental
representation. 

We should consider four cases:

\item{1.}  The anomaly \anomaly\ vanishes, $\alpha=c=0$.  In this case
the theory is consistent with a coupling constant $g$.  It multiplies an
irrelevant operator such that at long distance the theory is free.  There
could be an interesting fixed point at $g=\infty$.

\item{2.} The anomaly \anomaly\ does not vanish and $\alpha=0$, $c>0$.
In this case we can cancel the anomaly by adding a tensor multiplet.  We
will consider this case below.

\item{3.} The anomaly \anomaly\ does not vanish and $\alpha=0$, $c<0$.
In this case we cannot cancel the anomaly with a tensor multiplet.  The
only way to cancel it is by coupling the theory to gravity and to use
the two form with an anti-self-dual field strength in the gravity
multiplet to cancel the anomaly.  Therefore, in the absence of gravity,
such a field theory is inconsistent.  In other words, its gauge coupling
has to satisfy ${1 \over g^2} \sim M_{Planck}^2$.

\item{4.} The anomaly \anomaly\ does not vanish and $\alpha \not=0$.
These theories are anomalous and are inconsistent.  The anomaly cannot
be removed by adding more fields.

In the rest of this note we will focus on the second case.  As an
example consider $SU(2)$ gauge theories with $N_f$ hypermultiplets in
the fundamental representation.  The anomaly is easily computed
\eqn\sutwoan{ \half (16-N_f) \tr F^2 \tr F^2 .}

The anomaly $c=2(16-N_f)$ fits nicely with a pattern of similar
anomalies in $SU(2)$ gauge theories in $d=3,4,5$ dimensions \fived\
where $c=2(2^{d-2}-N_f)$.  They are all generated only at one loop, they
receive contributions only from small representations of supersymmetry.
Hypermultiplets and vector multiplets contribute with opposite
signs.  Also, as explained in \fived\ this quantity has a natural
interpretation in terms of the physics of branes in string theory
\ref\nclp{For a nice review see, S. Chaudhuri, C. Johnson, and J.
Polchinski, ``Notes on D-Branes,'' hep-th/9602052.}.

In all of these cases, the sign of $c$ is of crucial importance for the
behavior of the theory.  In four dimensions, $c$ coincides with the
coefficient of the one loop beta function.  Therefore, its sign
determines whether the theory becomes strongly coupled at short distance
or not.  It also determines whether non-perturbative effects can affect
the singularities on the moduli space of vacua.  In three dimensions the
theory is always asymptotically free but the sign of $c$ still
determines whether the singularities in the moduli space are modified or
not and whether instantons affect the metric on that space
\ref\threedtwo{N. Seiberg and E. Witten, ``Gauge Dynamics and
Compactification to Three Dimensions,'' hep-th/9607163.}.
In five dimensions the theories are always IR free.  However, the sign
of $c$ determines whether the strong coupling limit of the theories
exists.  For $c>0$ this limit leads to non-trivial fixed points \fived.

In six dimensions the sign of $c$ is similarly important.  For $N_f>16$,
$c<0$ and the theory is anomalous.  This means that there is no
consistent field theory.  For $N_f=16$, $c=0$ and the theory is anomaly
free.  In this theory the gauge coupling $g$ is a parameter.  This is
the theory on the 5-brane of small SO(32) instantons
\ref\witsmi{E. Witten, ``Small Instantons in String Theory,''
hep-th/9511030, \np{460}{1995}{541}.}.

For $N_f<16$, $c>0$ and the anomaly can be cancelled by adding a
tensor multiplet to the theory.  The scalar field in this multiplet
$\Phi$ leads to a new one real parameter family of vacua.  Although we
cannot write a Lorentz invariant Lagrangian for the two form in the
tensor multiplet, the interactions of $\Phi$ and the gauge bosons are
\eqn\bosint{{1 \over g^2} F_{\mu\nu}^2 +(\partial \Phi)^2 + \sqrt c \Phi
F_{\mu\nu}^2 }
where it is clear that $c$ should be positive.  The last term in
\bosint\ allows us absorb the gauge coupling $1 \over g^2$ in $\Phi$.
Then, the bosonic terms are
\eqn\bosinti{(\partial \Phi)^2 + \sqrt c \Phi F_{\mu\nu}^2 }
and the effective gauge coupling is ${1 \over g_{eff}^2(\Phi)}= \sqrt c
\Phi$.  For large $\Phi$ the gauge theory is weakly coupled but there is
a strong coupling point at $\Phi=0$.  At this point the theory could have
a non-trivial fixed point.  As a rather weak test of this proposal,
note that the terms in \bosinti\ (and also all the other terms in the
Lagrangian) are scale invariant.

It is easy to include $N_a$ hypermultiplets in the adjoint of $SU(2)$ in
this discussion.  In this case $c= 2(16-N_f-16N_a)$.  For $N_f=0$ and
$N_a=1$, $c$ vanishes.  This theory has (1,1) supersymmetry in six
dimension.  Since the anomaly vanishes, $g$ is a parameter in the theory
and it multiplies an irrelevant operator.  For all other cases (as well
as with higher dimensional representations of $SU(2)$) the anomaly $c$
is negative.  It is also straightforward to generalize this discussion
of $SU(2)$ to other groups and even to non-simple groups.  In some cases
more than one tensor multiplet is needed to cancel the anomaly.

\nref\mandf{E. Witten, ``Phase Transitions in M-Theory and F-Theory,'' 
hep-th/9603150.}%
\nref\vafaf{C. Vafa, ``Evidence for F-theory,'' hep-th/9602022.}%
\nref\vafamorI{D. Morrison and C. Vafa, ``Compactifications of F-Theory
on Calabi-Yau Three-Folds I,'' DUKE-TH-96-106, hep-th/9602114.}%
\nref\vafamorII{D. Morrison and C. Vafa, ``Compactifications of F-Theory
on Calabi-Yau Three-Folds II,'' DUKE-TH-96-107, hep-th/9603161.}%

The main question is which of these singularities indeed leads to a
non-trivial fixed point of the renormalization group.  We cannot give
a general answer to this question.  However, in many cases it is
known, using string ideas, that the corresponding fixed point exists.
These theories arise in compactifications of string theory to six
dimensions.  Many of these compactifications correspond to
compactifications of the $E_8\times E_8$ heterotic string on K3 with
instanton numbers $(n_1,n_2)$ with $n_1+n_2=24$ in the two $E_8$
factors.  More general values of $(n_1,n_2)$ appear in
compactifications of M theory \refs{\ganoha, \seiwit, \mandf} and F
theory \refs{\vafaf\ -- \vafamorII}.  The strong coupling
singularities in these compactifications were first pointed out in
\ref\dmw{M.J. Duff, R. Minasian and E. Witten, ``Evidence for
Heterotic/Heterotic Duality,'' CTP-TAMU-54/95, hep-th/9601036,
\np{465}{1996}{413}.} 
and were interpreted as associated with tensionless strings in
\refs{\seiwit, \dlp}.  In this framework the theories are based on the
gauge groups $G=E_8$ with
no hypermultiplets, $G=E_7$ with $n=0,...,8$ half hypermultiplets in
the representation $\bf 56$ or various subgroups of $E_7$ with matter
content obtained by the Higgs mechanism of these theories.  For
example, we can get $SU(2)$ with $N_f=4, 10, 16$.  In all these cases
the low energy theory is embedded in string theory and the singularity
leads to a non-trivial fixed point.  It is easy to check that all
these theories satisfy our criteria for a fixed point.
Another non-trivial fixed point without any gauge field is the theory of
the small $E_8$ instantons \refs{\ganoha,\seiwit}.  These fixed points
are also considered in
\ref\newwit{E. Witten, to appear.}.

We presented some necessary conditions for an interacting fixed point in
$N=1$ supersymmetry in six dimensions: $\alpha=0$ and $c>0$ in \anomaly.
We also demonstrated them in a number of examples based on string
theory. We conjecture that these two conditions ($\alpha=0$ and $c>0$ in
\anomaly) are also sufficient.  As a trivial consistency check of this
conjecture, note that by adding decoupled hypermultiplets or vector
multiplets all these theories can be coupled to gravity.  All
gravitational and mixed anomalies can then be cancelled using the two
form in the gravity multiplet.

\medskip
\centerline{\bf Acknowledgements}
This work was supported in part by DOE grant DE-FG02-96ER40559.  We
thank T. Banks, S. Shenker and E. Witten for helpful discussions.

\bigskip

\listrefs
\end